\documentclass{article}
\usepackage{arxiv}
\usepackage{newunicodechar}
\newunicodechar{Ö}{\"O}\newunicodechar{ü}{\"u}\newunicodechar{ı}{{\i}}
\newunicodechar{í}{\'{\i}}\newunicodechar{ř}{\v{r}}\newunicodechar{š}{\v{s}}\newunicodechar{ú}{\'u}
\usepackage{amssymb,mathtools,bbm,mathrsfs,aliascnt}
\usepackage{amsmath}
\usepackage{amsthm}
\usepackage[numbers,sort&compress]{natbib}
\providecommand{\textcite}{\citet}
\usepackage{graphicx}
\usepackage{hyperref}
\usepackage{cleveref}

\usepackage{aliascnt}
\theoremstyle{plain}
\newtheorem{theorem}{Theorem}[section]
\newaliascnt{corollary}{theorem}
\newtheorem{corollary}[corollary]{Corollary}
\aliascntresetthe{corollary}
\newaliascnt{lemma}{theorem}
\newtheorem{lemma}[lemma]{Lemma}
\aliascntresetthe{lemma}
\newaliascnt{proposition}{theorem}
\newtheorem{proposition}[proposition]{Proposition}
\aliascntresetthe{proposition}
\theoremstyle{definition}
\newaliascnt{definition}{theorem}
\newtheorem{definition}[definition]{Definition}
\aliascntresetthe{definition}
\newaliascnt{problem}{theorem}

\aliascntresetthe{problem}
\newaliascnt{example}{theorem}

\aliascntresetthe{example}
\theoremstyle{remark}
\newaliascnt{remark}{theorem}
\newtheorem{remark}[remark]{Remark}
\aliascntresetthe{remark}
\newaliascnt{fact}{theorem}

\aliascntresetthe{fact}
\crefname{problem}{Problem}{Problems}
\Crefname{problem}{Problem}{Problems}
\crefname{fact}{Fact}{Facts}
\Crefname{fact}{Fact}{Facts}
\input{preamble.sty}

\title{Exact signal recovery from generative priors with sigmoid output activation}
\author{\textbf{Matthew S. Scott}\thanks{Corresponding author.}\\\texttt{matthewscott@math.ubc.ca} \And \textbf{Yaniv Plan}\\\texttt{yaniv@math.ubc.ca} \And \textbf{Özgür Yılmaz}\\\texttt{oyilmaz@math.ubc.ca} \AND Department of Mathematics, University of British Columbia, Vancouver, BC, Canada}
\begin{document}
\maketitle
\begin{abstract}
We show exact signal recovery in generative compressed sensing for feedforward neural networks with piecewise-linear activations and a final sigmoid activation, with no assumption of expansivity. Consider a ReLU network of depth $d$ with a sigmoid final activation, whose input layer, of dimension $k$, is the narrowest, and let $K$ be the geometric mean of the widths of its layers. We show that, with high probability, all signals of the range whose entries lie in $[\delta, 1 - \delta]$ are uniformly recovered from $m \gtrsim k\big(d \log(K/k) + \log(1+\log(1/\delta))\big)$ measurements.
\end{abstract}
\section{Introduction}
\label{loc:body.introduction}
Compressed sensing recovers a signal $x_0 \in \mathbb{R}^n$ from $m \ll n$ linear measurements $y = Ax_0 + \xi$, with deterministic noise $\xi \in \mathbb{R}^m$, by leveraging a prior assumption on the structure of the true signal. Classical theory uses sparsity. Generative compressed sensing uses a trained generator instead: natural images lie close to the range of a network $G:\mathbb{R}^k \to \mathbb{R}^n$ trained on them, and searching that range for the image whose measurements best match $y$ recovers images from far fewer measurements than sparsity needs \cite{boraCompressedSensingUsing2017}.

How many measurements suffice is a question about the complexity of the range: the measurement matrix must not collapse the difference between two images of the range, so that they may be differentiated from measurements alone. A random matrix with few rows preserves the norm of any one fixed vector with high probability; the difficulty is to do so for all differences of points in the range at once. The more complex the range, the more directions these differences take, and the more likely a random matrix is to collapse some of them.

The complexity of the range of feedforward neural networks has been quantified in several ways. For networks with rectified linear units (ReLU), the range is a union of $N$ pieces of dimension at most $k$. A union bound over pairs of pieces shows that, once $m \ge C(k + \log N)$ for an absolute constant $C$, $A$ has the restricted isometry property (RIP) on the differences of points in the range: it preserves their lengths up to a constant factor, and recovery of signals in the range from noiseless measurements is exact \cite{boraCompressedSensingUsing2017}, \cite{naderiIndependentMeasurementsGeneral2021}. Here and in the rest of the paper, $C, C_0, c_0$ denote absolute constants, which may differ from one appearance to the next.

For general $L$-Lipschitz networks with inputs in a ball of radius $r$, \textcite{boraCompressedSensingUsing2017} obtain recovery up to an accuracy of $\eta$ from $m \ge Ck\log(Lr/\eta)$ measurements. The generality of the Lipschitz route has a cost: the Lipschitz constant of a trained network is typically very large, and the sample complexity it gives is then often an artifact of the theory rather than a property of the range. For example, for a linear generator $G(z) = Wz$ with $W \in \mathbb{R}^{n \times k}$ of rank $k$, the range is a $k$-dimensional subspace, so $m = Ck$ measurements suffice, while $L = \|W\|$ is arbitrary. The sample complexity is governed by the geometry of the range, not by the Lipschitz constant. Piece counting captures this geometry for piecewise-linear networks, but decoders commonly end in a sigmoid, and their range has no flat pieces.

\textcite{gajjarSubspaceEmbeddingsNonlinear2020} use piecewise-linear approximations to obtain embeddings for subspaces under smooth entrywise nonlinearities. Their pairwise-distance theorem covers generators of the form $\sigma(Wz)$, without bounds on the input or the weight matrix, but includes an additive error term. They also prove purely relative norm embeddings for certain nonlinearities; however, these do not directly provide the \emph{secant} estimates required for signal recovery. In the many-layer case, they use the Lipschitz constant of the network. We also approximate the sigmoid with a piecewise-linear function, but in relative error, enabling exact recovery in the noiseless setting. Our results do not depend on the Lipschitz constant of the network, and our sample complexity matches that of piecewise-linear networks of the same architecture, up to an additive doubly-logarithmic term.

For ReLU networks whose input layer is the narrowest, we provide a signal recovery result for a final sigmoid activation, uniform over the signals away from saturation.
\begin{theorem}[Simplified version of \Cref{loc:gaussian_recovery_of_unsaturated_signals_without_truncation.statement}]
\label{loc:gaussian_recovery_of_unsaturated_signals_without_truncation.simplified_statement}
Let $G\colon \mathbb{R}^k \to \mathbb{R}^n$ be a feedforward network of depth $d$ with ReLU intermediate activations, sigmoid final activation and widths $k = k_0, k_1, \dots, k_{d-1}$ in $\mathbb{N}$, let $\delta \in (0, \frac12]$ be a saturation level, and let $A \in \mathbb{R}^{m \times n}$ be a matrix with independent $N(0, 1/m)$ entries. With $k_d := n$, suppose that the input layer is the narrowest, $k \le k_i$ for $1 \le i \le d$, and let $K := \big(\prod_{i=1}^{d} k_i\big)^{1/d}$ be the geometric mean of the widths of the layers of the network. We require that $K \ge e\,k$. There are absolute constants $C, c_0 > 0$ such that if
\begin{equation*}
m \ge C\,k\Big(d\,\log\frac{K}{k} + \log\Big(1 + \log\frac1\delta\Big)\Big),
\end{equation*}
then with probability at least $1 - 3e^{-c_0 m}$, for every signal $x_0 \in \operatorname{range}(G)$ whose entries lie in $[\delta, 1 - \delta]$, noise $\xi \in \mathbb{R}^m$ and tolerance $\varepsilon \ge 0$, with measurements $y := Ax_0 + \xi \in \mathbb{R}^m$, every $\hat x \in \operatorname{range}(G)$ such that $\|A\hat x - y\|_2^2 \le \inf_{x \in \operatorname{range}(G)}\|Ax - y\|_2^2 + \varepsilon$ satisfies
\begin{equation*}
\|\hat x - x_0\|_2 \le 4\|\xi\|_2 + 2\sqrt\varepsilon .
\end{equation*}
\end{theorem}
This result follows directly from \Cref{loc:gaussian_recovery_of_unsaturated_signals_without_truncation.statement}; we provide the details in the appendix.

Our method is to first prove a sharper bound on the number of linear pieces in the range of a piecewise-linear feedforward network, improving on \cite[Lemma 3.1]{naderiIndependentMeasurementsGeneral2021}: we truncate the network at its narrowest layer, and the bound holds for any piecewise-linear activation with a finite hyperplane cover, with biases (\Cref{loc:body.complexity_of_piecewise:linear_feedforward_neural_networks}). Counting the linear pieces of a network is also a classical question in the study of expressivity \cite{montufarNotesNumberLinear2017}, \cite{serraBoundingCountingLinear2018}, \cite{hinzFrameworkConstructionUpper2019}. Compared with these counts, we obtain a tighter bound by recognizing that all pieces originating from layers preceding the bottleneck layer get merged together at the bottleneck layer, and therefore do not contribute to the complexity of the range. This improvement is only meaningful when considering the number of pieces in the range; if one cares about the number of pieces of $G$ as a piecewise-linear function, then earlier layers cannot be ignored.

We then approximate the network in a metric suited to signal recovery. Consider $\sigma \circ H$, a smooth output activation $\sigma$ after a piecewise-linear network $H$, and $\bar\sigma \circ H$ for a piecewise-linear approximation $\bar\sigma$ of $\sigma$. The network $\bar\sigma \circ H$ is piecewise linear, and by counting its pieces we can guarantee the RIP on the difference vectors of its range with high probability. The RIP depends only on the directions of the difference vectors, not their lengths, so the RIP of $\bar\sigma \circ H$ passes to $\sigma \circ H$ when each difference vector of $\sigma \circ H$ is close in angle to some difference vector of $\bar\sigma \circ H$. A sufficient condition is that, for some small $c$ and all $z, z' \in \mathbb{R}^k$,
\begin{equation*}
\|(\sigma(H(z'))-\sigma(H(z))) - (\bar\sigma(H(z'))-\bar\sigma(H(z)))\|_2 \le c\,\|\sigma(H(z'))-\sigma(H(z))\|_2.
\end{equation*}
Requiring the approximation $\bar\sigma$ to be uniformly close to $\sigma$ would be the wrong metric: it leaves an additive error and fails to guarantee exact recovery.
Instead, we choose $\bar\sigma$ to approximate $\sigma$ relatively: for all real $s, t$,
\begin{equation*}
|(\sigma(t)-\sigma(s)) - (\bar\sigma(t)-\bar\sigma(s))| \le c\,|\sigma(t)-\sigma(s)| ,
\end{equation*}
which gives the bound above coordinate by coordinate, since the two networks share $H$. This works for any smooth output activation that admits such an approximation; we construct one for the sigmoid, a common choice in practice; other smooth output activations can be treated similarly. The main difficulty is that, for the sigmoid, such a piecewise-linear approximation $\bar\sigma$ must have infinitely many pieces. We get around this by restricting the uniform guarantee to the signals away from saturation, whose entries lie in $[\delta, 1-\delta]$ for a saturation level $\delta$ fixed in advance. Their pre-activations lie in a bounded window, and we only need the bound above for $s$ in that window, which one approximation with finitely many pieces achieves for all $t$ (\Cref{loc:body.smooth_output_activation}). The sample complexity then depends, through a double logarithm, on how far into saturation the sigmoid may go to generate the true signal: the saturation level adds only a term $k\log(1+\log(1/\delta))$, about $4k$ for $\delta = 10^{-30}$.
\section{Complexity of piecewise-linear feedforward neural networks}
\label{loc:body.complexity_of_piecewise:linear_feedforward_neural_networks}
The standard activation is the ReLU $t \mapsto \max\{t, 0\}$ on $\mathbb{R}$, applied entrywise to $x \in \mathbb{R}^p$. Its only \emph{kink}, the point where it is not differentiable, is $0$. The kinks of the entries induce hyperplanes in the domain $\mathbb{R}^p$ of the activation, namely $\{x \in \mathbb{R}^p : x_\nu = 0\}$, $1 \le \nu \le p$. Such a finite set of affine hyperplanes is called an \emph{arrangement} of hyperplanes. An arrangement splits the space into \emph{cells}, the connected components between the hyperplanes. The activation is affine wherever no entry of $x$ changes sign, that is, on each cell of the arrangement of these $p$ hyperplanes. The same holds for any continuous piecewise-linear entrywise activation, with one hyperplane per entry and kink. Activations that are not entrywise, such as max-pooling, are still affine on the cells of finitely many hyperplanes, and the number of these hyperplanes is what our bounds use.
\begin{definition}
\label{loc:hyperplane_cover_of_a_piecewise:linear_map.statement}
A \emph{hyperplane cover} of a continuous piecewise-linear map $\phi:\mathbb{R}^p \to \mathbb{R}^p$ is a finite set of affine hyperplanes in the domain $\mathbb{R}^p$ of $\phi$ such that $\phi$ is affine on each cell of their arrangement.
\end{definition}
\begin{definition}
\label{loc:feedforward_network_with_piecewise:linear_activations.statement}
A \emph{feedforward network} $G:\mathbb{R}^k \to \mathbb{R}^n$ with $d$ layers of widths $k = k_0, k_1, \dots, k_d = n$ in $\mathbb{N}$ is
\begin{equation*}
G = \phi_d \circ A_d \circ \cdots \circ \phi_1 \circ A_1,
\end{equation*}
with affine maps $A_i:\mathbb{R}^{k_{i-1}} \to \mathbb{R}^{k_i}$, $A_i(x) = W^{(i)}x + \beta^{(i)}$, where $W^{(i)} \in \mathbb{R}^{k_i \times k_{i-1}}$, $\beta^{(i)} \in \mathbb{R}^{k_i}$, and \emph{activations} $\phi_i:\mathbb{R}^{k_i} \to \mathbb{R}^{k_i}$, $1 \le i \le d$, which are continuous piecewise-linear maps, each with a finite \hyperref[loc:hyperplane_cover_of_a_piecewise:linear_map.statement]{hyperplane cover}.

The parameters of $G$ are the \emph{depth} $d$, the \emph{bottleneck layer} $b := \max\{i \le d : k_i = \min_{0 \le i' \le d} k_{i'}\}$, and the \emph{hyperplane counts} $M_i$, $1 \le i \le d$, the least number of hyperplanes in a hyperplane cover of $\phi_i$.
\end{definition}
\begin{remark}
\label{loc:feedforward_network_with_piecewise:linear_activations.max:pooling}
For a max-pooling activation, fix a \emph{pool size} $p \in \mathbb{N}$ dividing $k_i$, and split the coordinates of $x \in \mathbb{R}^{k_i}$ into $k_i/p$ \emph{blocks} of $p$ consecutive coordinates, $(x_1, \dots, x_p), (x_{p+1}, \dots, x_{2p}), \dots$. A max-pooling activation returns the maximum of each block, padded with zeros to keep the width $k_i$. The maximum of a block is linear on each set where a given coordinate of the block is largest, and these sets are cut out by the hyperplanes $\{x \in \mathbb{R}^{k_i} : x_\nu = x_{\nu'}\}$, one for each pair of indices $\nu \ne \nu'$ in a common block. Each block contributes $\binom{p}{2}$ hyperplanes, so $M_i \le \frac{k_i}{p}\binom{p}{2} = \frac{(p-1)k_i}{2}$. A similar hyperplane count for max-pooling appears in \cite[Theorem 10]{serraBoundingCountingLinear2018}.
\end{remark}
\begin{lemma}
\label{loc:cells_of_a_hyperplane_arrangement.statement}
Consider an arrangement of $M \in \{0\} \cup \mathbb{N}$ affine hyperplanes in $\mathbb{R}^D$, $D \in \mathbb{N}$. It has at most
\begin{equation*}
\Big(1 + \frac{eM}{D}\Big)^D
\end{equation*}
cells.
\end{lemma}
\begin{proof}[\hypertarget{loc:cells_of_a_hyperplane_arrangement.proof}Proof of \Cref{loc:cells_of_a_hyperplane_arrangement.statement}]\phantomsection\label{loc:cells_of_a_hyperplane_arrangement.proof}

By \cite[Proposition 6.1.1]{matousekLecturesDiscreteGeometry2013}, the cells number at most $\sum_{i=0}^{D}\binom{M}{i}$. If $M < D$, then $\binom{M}{i} = 0$ for $i > M$, so this sum is $\sum_{i=0}^{M}\binom{M}{i} = 2^M \le (1 + e)^M \le (1 + eM/D)^D$, as $x \mapsto (1 + eM/x)^x$ is nondecreasing on $x > 0$ and equals $(1 + e)^M$ at $x = M$. If $M \ge D$, then $\binom{M}{i} \le M^i/i!$ and $M/D \ge 1$ give
\begin{equation*}
\sum_{i=0}^{D}\binom{M}{i} \le \sum_{i=0}^{D}\frac{M^i}{i!} = \sum_{i=0}^{D}\frac{D^i}{i!}\Big(\frac{M}{D}\Big)^i \le \Big(\frac{M}{D}\Big)^D\sum_{i=0}^{D}\frac{D^i}{i!} \le \Big(\frac{eM}{D}\Big)^D \le \Big(1 + \frac{eM}{D}\Big)^D.
\end{equation*}
\end{proof}
\begin{lemma}
\label{loc:image_of_a_subspace_cover_under_a_piecewise:linear_map.statement}
Let $p, D, N \in \mathbb{N}$, let $E_1, \dots, E_N \subseteq \mathbb{R}^p$ be affine subspaces, each of dimension at most $D$, and let $\phi:\mathbb{R}^p \to \mathbb{R}^p$ be continuous with a hyperplane cover, as in \Cref{loc:hyperplane_cover_of_a_piecewise:linear_map.statement}, of $M \in \{0\} \cup \mathbb{N}$ hyperplanes. Then $\phi(E_1 \cup \dots \cup E_N)$ lies in a union of at most $N\,(1 + eM/D)^D$ affine subspaces of $\mathbb{R}^p$, each of dimension at most $D$.
\end{lemma}
\begin{proof}[\hypertarget{loc:image_of_a_subspace_cover_under_a_piecewise:linear_map.proof}Proof of \Cref{loc:image_of_a_subspace_cover_under_a_piecewise:linear_map.statement}]\phantomsection\label{loc:image_of_a_subspace_cover_under_a_piecewise:linear_map.proof}

It suffices to show that each $\phi(E_l)$, $l \in \{1, \dots, N\}$, lies in a union of at most $(1 + eM/D)^D$ affine subspaces of dimension at most $D$.

Fix $l$, write $E := E_l$, let $D' \le D$ be its dimension, and let $\mathcal H$ be the hyperplane cover of $\phi$, with $M$ hyperplanes. At most $M$ hyperplanes cut $E \cong \mathbb{R}^{D'}$ in a hyperplane of $E$. If $D' = 0$, $E$ is a point and a single cell. Otherwise, by \Cref{loc:cells_of_a_hyperplane_arrangement.statement}, $E$ has at most $(1 + eM/D')^{D'} \le (1 + eM/D)^D$ \emph{cells}, whose closures cover $E$; the inequality holds as $x \mapsto (1 + eM/x)^x$ is nondecreasing on $x > 0$.

We show that $\phi$ is affine on the closure $\bar V$ of each cell $V$ of $E$, by showing that $V$ lies in the closure of a single cell $C$ of $\mathbb{R}^p$. The cells of $\mathbb{R}^p$ are described by picking a side for each hyperplane in the arrangement: one of its two open half-spaces. The points lying in every picked half-space form a cell, when that set is nonempty. Fix a direction $v \in \mathbb{R}^p$ parallel to no hyperplane of $\mathcal H$. For each $H \in \mathcal H$, we have either:

\begin{itemize}
\item $H$ does not contain $E$, and then $V$, being connected and disjoint from $H$, lies in one open half-space of $H$; we pick that one; or
\item $H$ contains $E$, and then $V \subseteq H$ lies on neither side; we pick the side towards which $v$ points.
\end{itemize}

Let $C$ be the set of points lying in every picked half-space. It is nonempty: pick any $y \in V$. For each hyperplane containing $E$, the point $y + tv$ lies on the side towards which $v$ points for every $t > 0$, as $y$ lies on the hyperplane; for each other hyperplane, $y$ lies strictly on the picked side, and so does $y + tv$ for small enough $t > 0$, by continuity. So $C$ is a cell of $\mathbb{R}^p$ whose closure $\bar C$ contains $V$. The map $\phi$ is affine on $C$, by definition of a hyperplane cover, therefore on $\bar C$, by continuity, and therefore on $\bar V$.

Thus each $\phi(\bar V)$ lies in an affine subspace of dimension at most $D$, and $\phi(E) = \bigcup_{V} \phi(\bar V)$ is covered by at most $(1 + eM/D)^D$ such subspaces.
\end{proof}
The next proposition covers the range by few low-dimensional pieces. By truncating the network at its narrowest layer, it improves on the piece count of \cite[Lemma 3.1]{naderiIndependentMeasurementsGeneral2021} and requires no expansivity.
\begin{proposition}
\label{loc:range_cover_of_a_piecewise:linear_network.statement}
Let $G$ be a network as in \Cref{loc:feedforward_network_with_piecewise:linear_activations.statement}, with parameters $b$, $d$, $M_1, \dots, M_d$. Then there are affine subspaces $E_1, \dots, E_N \subseteq \mathbb{R}^n$, each of dimension at most $k_b$, with
\begin{equation*}
\operatorname{range}(G) \subseteq \bigcup_{l=1}^{N} E_l, \qquad N \le \prod_{j=b+1}^{d}\Big(1 + \frac{e M_j}{k_b}\Big)^{k_b}.
\end{equation*}
\end{proposition}
\begin{remark}
\label{loc:range_cover_of_a_piecewise:linear_network.low:rank_weights}
This bound can be further refined by using the minimal rank of the weight matrices in place of the minimal width.
\end{remark}
\begin{proof}[\hypertarget{loc:range_cover_of_a_piecewise:linear_network.proof}Proof of \Cref{loc:range_cover_of_a_piecewise:linear_network.statement}]\phantomsection\label{loc:range_cover_of_a_piecewise:linear_network.proof}

Write $\Phi(M) := (1 + eM/k_b)^{k_b}$.

If $b = d$, then $k_b = n$, and the claim holds with $N = 1$ and $E_1 := \mathbb{R}^n$. Let $b < d$ from now on.

For $b < i \le d$, define the network from the bottleneck to layer $i$ to be
\begin{equation*}
G^{(i)} := \phi_i \circ A_i \circ \cdots \circ \phi_{b+1} \circ A_{b+1} \colon \mathbb{R}^{k_b} \to \mathbb{R}^{k_i},
\end{equation*}
and let $G^{(b)}$ be the identity map of $\mathbb{R}^{k_b}$. The network after the bottleneck is $G^{(d)} \colon \mathbb{R}^{k_b} \to \mathbb{R}^n$, so that $G = G^{(d)} \circ \phi_b \circ A_b \circ \cdots \circ \phi_1 \circ A_1$, and $G = G^{(d)}$ if $b = 0$. Hence $\operatorname{range}(G)$ lies in the image $G^{(d)}(\mathbb{R}^{k_b})$. We argue by induction that for every $i \in \{b, \dots, d\}$, the image $G^{(i)}(\mathbb{R}^{k_b})$ lies in a union of $N_i$ affine subspaces of $\mathbb{R}^{k_i}$, each of dimension at most $k_b$, with
\begin{equation}
\label{eq:piece:count}
N_i \le \prod_{j=b+1}^{i}\Phi(M_j).
\end{equation}
For $i = b$, take the single subspace $\mathbb{R}^{k_b}$, so $N_b = 1$. For $i > b$, let $F_1, \dots, F_{N_{i-1}} \subseteq \mathbb{R}^{k_{i-1}}$ be the affine subspaces given by induction. Since $A_i$ is affine, each $A_i(F_l)$ is an affine subspace of $\mathbb{R}^{k_i}$ of dimension at most $k_b$, and their union contains $A_i(G^{(i-1)}(\mathbb{R}^{k_b}))$. The activation $\phi_i \colon \mathbb{R}^{k_i} \to \mathbb{R}^{k_i}$ is continuous with a hyperplane cover of $M_i$ hyperplanes, so by \Cref{loc:image_of_a_subspace_cover_under_a_piecewise:linear_map.statement}, with $p := k_i$ and $D := k_b$, the image $G^{(i)}(\mathbb{R}^{k_b}) = \phi_i(A_i(G^{(i-1)}(\mathbb{R}^{k_b})))$ lies in a union of at most $N_{i-1}\,\Phi(M_i)$ affine subspaces of $\mathbb{R}^{k_i}$, each of dimension at most $k_b$. Hence \Cref{eq:piece:count} holds for $i$.

At $i = d$, as $k_d = n$, this gives affine subspaces $E_1, \dots, E_N \subseteq \mathbb{R}^n$, each of dimension at most $k_b$, with $N \le \prod_{j=b+1}^{d}\Phi(M_j)$ and
\begin{equation*}
\operatorname{range}(G) \subseteq G^{(d)}(\mathbb{R}^{k_b}) \subseteq \bigcup_{l=1}^{N} E_l.
\end{equation*}
\end{proof}
The bound is increasing in each $M_j$, so an upper bound on the number of hyperplanes may be substituted for it.
\begin{definition}
\label{loc:restricted_isometry_property_with_a_constant.statement}
Let $\kappa \in (0, 1)$. A matrix $A \in \mathbb{R}^{m \times n}$ satisfies the $\kappa$-\emph{RIP} (restricted isometry property with constant $\kappa$) on a set $S \subseteq \mathbb{R}^n$ if
\begin{equation*}
\big|\|Ah\|_2 - \|h\|_2\big| \le \kappa\|h\|_2 \qquad \text{for all } h \in S .
\end{equation*}
\end{definition}
\begin{proposition}
\label{loc:rip_on_the_range_of_a_piecewise:linear_network.statement}
Let $G$ be a network as in \Cref{loc:feedforward_network_with_piecewise:linear_activations.statement}, with parameters $b, d, M_1, \dots, M_d$, and $N$ such that
\begin{equation*}
\log N = k_b\sum_{j=b+1}^{d}\log\Big(1 + \frac{e\,M_j}{k_b}\Big).
\end{equation*}
If $A \in \mathbb{R}^{m \times n}$ has independent $N(0, 1/m)$ entries and $m \ge C_0(k_b + \log N)$, then with probability at least $1 - 2e^{-c_0 m}$, $A$ satisfies the $\frac13$-RIP on $\operatorname{range}(G) - \operatorname{range}(G)$; $C_0, c_0$ are absolute constants.
\end{proposition}
\begin{proof}[\hypertarget{loc:rip_on_the_range_of_a_piecewise:linear_network.proof}Proof of \Cref{loc:rip_on_the_range_of_a_piecewise:linear_network.statement}]\phantomsection\label{loc:rip_on_the_range_of_a_piecewise:linear_network.proof}

Let $E_1, \dots, E_{N'}$, with $N' \le N$, be the affine subspaces of \Cref{loc:range_cover_of_a_piecewise:linear_network.statement}. A chord $G(z') - G(z)$ lies in some $E_i - E_{i'}$, which spans a linear subspace of dimension at most $2k_b + 1$, so the chords of $G$ lie in the union $\mathcal T$ of the at most $N^2$ subspaces $\operatorname{span}(E_i - E_{i'})$. On a fixed subspace of dimension $D$, a Gaussian matrix satisfies $\frac23\|t\| \le \|At\| \le \frac43\|t\|$ for all $t$ in it with probability at least $1 - 2e^{-c_0 m}$ once $m \ge C_0D$ (\cite[Theorem 4.6.1]{vershyninHighDimensionalProbabilityIntroduction2018}). A union bound over the subspaces of $\mathcal T$ gives these bounds on all of $\mathcal T$, hence on the chords, the factor $N^2$ being absorbed, for $C_0$ large enough, since then $2\log N \le c_0 m/2$, at the cost of halving $c_0$, as in \cite{boraCompressedSensingUsing2017} and \cite{naderiIndependentMeasurementsGeneral2021}.
\end{proof}
\section{Smooth output activation}
\label{loc:body.smooth_output_activation}
We now consider networks $G$ whose final activation is the sigmoid $\sigma$, all earlier activations being piecewise linear. The idea is to replace $\sigma$ by a piecewise-linear approximation $\bar\sigma$: the resulting network $\bar G$ is piecewise linear, so the proposition above gives the restricted isometry on its range, and the next lemma carries it over to $G$ when $\bar G$ is close to $G$. Closeness on the chords from every true signal $x_0$ of the range would need infinitely many pieces in $\bar\sigma$, since entries of $x_0$ near $0$ or $1$ come from arbitrarily large pre-activations. Instead, we fix a saturation level $\delta$ before $A$ is drawn, and only consider true signals $x_0$ whose entries lie in $[\delta, 1 - \delta]$. Their pre-activations lie in a bounded window, where a $\bar\sigma$ with few pieces suffices.
\begin{lemma}
\label{loc:transfer_of_the_restricted_isometry_between_two_networks.statement}
Let $A \in \mathbb{R}^{m \times n}$, let $0 \le c \le 1/(8 + 6\|A\|_{\mathrm{op}})$, and let $h, \bar h \in \mathbb{R}^n$. If $\|h - \bar h\|_2 \le c\,\|h\|_2$, then
\begin{equation*}
\tfrac23\|\bar h\|_2 \le \|A\bar h\|_2 \le \tfrac43\|\bar h\|_2 \quad\Longrightarrow\quad \tfrac12\|h\|_2 \le \|Ah\|_2 \le \tfrac32\|h\|_2.
\end{equation*}
\end{lemma}
\begin{proof}[\hypertarget{loc:transfer_of_the_restricted_isometry_between_two_networks.proof}Proof of \Cref{loc:transfer_of_the_restricted_isometry_between_two_networks.statement}]\phantomsection\label{loc:transfer_of_the_restricted_isometry_between_two_networks.proof}

The first hypothesis gives $(1 - c)\|h\| \le \|\bar h\| \le (1 + c)\|h\|$ and $\|A(h - \bar h)\| \le c\,\|A\|_{\mathrm{op}}\|h\|$. By the triangle inequality and the second,
\begin{equation*}
\|Ah\| \le \|A\bar h\| + \|A(h - \bar h)\| \le \tfrac43(1 + c)\|h\| + c\|A\|_{\mathrm{op}}\|h\| \le \tfrac32\|h\|,
\end{equation*}
\begin{equation*}
\|Ah\| \ge \|A\bar h\| - \|A(h - \bar h)\| \ge \tfrac23(1 - c)\|h\| - c\|A\|_{\mathrm{op}}\|h\| \ge \tfrac12\|h\|,
\end{equation*}
since $c(\frac43 + \|A\|_{\mathrm{op}}) \le \frac16$.
\end{proof}
Applied at every chord from a signal away from saturation, the lemma passes the $\frac13$-RIP on the chords of $\bar G$ to the $\frac12$-RIP on those of $G$. For a Gaussian matrix $A$, the relative accuracy $c$ it asks for is of order $\sqrt{m/n}$.
\begin{remark}
\label{loc:transfer_of_the_restricted_isometry_between_two_networks.relative_closeness_is_a_condition_on_angles}
Write $\hat h := h/\|h\|_2$ and $\hat{\bar h} := \bar h/\|\bar h\|_2$ for the normalized chords. The restricted isometry of a chord depends only on its direction: the ratio $\|Ah\|_2/\|h\|_2$ is in fact $\|A\hat h\|_2$. So comparing $h$ with $\bar h$ means comparing $\|A\hat h\|_2$ with $\|A\hat{\bar h}\|_2$. These are close when the unit vectors $\hat h$ and $\hat{\bar h}$ are close, that is, when the angle $\theta$ between $h$ and $\bar h$ is small. The hypothesis of the statement makes this angle small:
\begin{equation*}
\|h - \bar h\|_2 \le c\,\|h\|_2 \quad\Longrightarrow\quad \sin\theta \le c.
\end{equation*}
\end{remark}
\subsection{Approximating the sigmoid}
\label{loc:body.smooth_output_activation.approximating_the_sigmoid}
It remains to build a suitable $\bar\sigma$: a piecewise-linear activation with few pieces such that $\bar G$ approximates $G$ well enough for the restricted isometry to transfer from $\bar G$ to $G$, as in \Cref{loc:transfer_of_the_restricted_isometry_between_two_networks.statement}. We begin with an approximation on a bounded interval.
\begin{lemma}
\label{loc:piecewise:linear_approximation_of_an_increasing_smooth_function.statement}
Let $\alpha < \beta$ be real numbers, $I := [\alpha, \beta]$, and let $f \in C^2(I)$ with $f' > 0$, let $V := \mathrm{TV}_I(\log f')$ and let $c > 0$. There is a continuous piecewise-linear $\bar f\colon I \to \mathbb{R}$ with $\bar f(\alpha) = f(\alpha)$, $\bar f(\beta) = f(\beta)$ and at most $\max\big(1, \lceil V/\log(1 + c) \rceil\big)$ pieces such that for all $u, v \in I$
\begin{equation*}
\big|(f - \bar f)(v) - (f - \bar f)(u)\big| \le c\,|f(v) - f(u)|.
\end{equation*}
\end{lemma}
\begin{proof}[\hypertarget{loc:piecewise:linear_approximation_of_an_increasing_smooth_function.proof}Proof of \Cref{loc:piecewise:linear_approximation_of_an_increasing_smooth_function.statement}]\phantomsection\label{loc:piecewise:linear_approximation_of_an_increasing_smooth_function.proof}

Let $\varepsilon := \log(1 + c)$, so that $e^{\varepsilon} - 1 = c$. Since $\mathrm{TV}_{[\alpha,t]}(\log f')$ is continuous in $t$, we can take nodes $\alpha = \tau_0 < \dots < \tau_J = \beta$ with $J \le \max\big(1, \lceil V/\varepsilon \rceil\big)$ and $\mathrm{TV}(\log f') \le \varepsilon$ on each subinterval $[\tau_{j-1}, \tau_j]$, and let $\bar f$ interpolate $f$. On a subinterval, $|\log f'(t) - \log f'(\rho)| \le \varepsilon$ for all $t, \rho$ in it, so $e^{-\varepsilon} f'(t) \le f'(\rho) \le e^{\varepsilon} f'(t)$. On each subinterval, the slope $\bar f'$ is the average of $f'$ over it, so it lies between the least and greatest value of $f'$ there, and $e^{-\varepsilon} f'(t) \le \bar f'(t) \le e^{\varepsilon} f'(t)$. Subtracting $f'(t)$ gives $(e^{-\varepsilon} - 1) f'(t) \le \bar f'(t) - f'(t) \le (e^{\varepsilon} - 1) f'(t)$, and since $1 - e^{-\varepsilon} \le e^{\varepsilon} - 1$, $|f'(t) - \bar f'(t)| \le (e^{\varepsilon} - 1) f'(t)$. Hence for $u < v$ in $I$,
\begin{equation*}
\big|(f - \bar f)(v) - (f - \bar f)(u)\big| \le \int_u^v |f'(t) - \bar f'(t)|\,dt \le (e^{\varepsilon} - 1)\int_u^v f'(t)\,dt = c\,(f(v) - f(u)).
\end{equation*}
\end{proof}
The transfer needs more: the approximation must follow the increments of the sigmoid from the pre-activation of every signal away from saturation, which lies in a bounded interval, to any other point of the line.
\begin{lemma}
\label{loc:fixed:grid_approximation_of_the_sigmoid.statement}
Let $\sigma\colon \mathbb{R} \to \mathbb{R}$ be the sigmoid, $\sigma(t) = 1/(1 + e^{-t})$, let $T \in [0, \infty)$ and $c \in (0, 1)$, and let $\tau := T + \log\frac1c + 3\log 2$. The \emph{piecewise-linear approximation} $\bar\sigma_T \colon \mathbb{R} \to \mathbb{R}$ is, on $[-\tau, \tau]$, the one of $\sigma$ given by \Cref{loc:piecewise:linear_approximation_of_an_increasing_smooth_function.statement} with accuracy parameter $c/4$, and is constant on the tails, $\bar\sigma_T = \sigma(-\tau)$ on $(-\infty, -\tau]$ and $\bar\sigma_T = \sigma(\tau)$ on $[\tau, \infty)$. It is continuous, has at most $\frac{10}{c}\big(T + \log\frac1c + 3\log 2\big) + 3$ pieces, and follows the increments of $\sigma$ from every point of $[-T, T]$ to relative accuracy $c$: for all $u \in [-T, T]$ and $v \in \mathbb{R}$,
\begin{equation*}
|(\sigma - \bar\sigma_T)(v) - (\sigma - \bar\sigma_T)(u)| \le c\,|\sigma(v) - \sigma(u)| .
\end{equation*}
\end{lemma}
\begin{proof}[\hypertarget{loc:fixed:grid_approximation_of_the_sigmoid.proof}Proof of \Cref{loc:fixed:grid_approximation_of_the_sigmoid.statement}]\phantomsection\label{loc:fixed:grid_approximation_of_the_sigmoid.proof}

We first count the pieces. As $|(\log\sigma')'| = |1 - 2\sigma| \le 1$, $\mathrm{TV}_{[-\tau,\tau]}(\log\sigma') \le 2\tau$, so \Cref{loc:piecewise:linear_approximation_of_an_increasing_smooth_function.statement} gives at most $\frac{2\tau}{\log(1 + c/4)} + 1$ pieces on $[-\tau, \tau]$, plus two tails, and $\log(1 + c/4) \ge \frac{c/4}{1 + c/4} \ge \frac{c}{5}$ as $c < 1$; $\bar\sigma_T$ is continuous because it agrees with $\sigma$ at $-\tau$ and $\tau$.

For the accuracy bound, write $\Delta := \sigma - \bar\sigma_T$, so that $\Delta(-\tau) = \Delta(\tau) = 0$, and let $u \in [-T, T] \subseteq [-\tau, \tau]$. If $v \in [-\tau, \tau]$, \Cref{loc:piecewise:linear_approximation_of_an_increasing_smooth_function.statement} gives $|\Delta(v) - \Delta(u)| \le \frac{c}{4}|\sigma(v) - \sigma(u)|$.

If $v < -\tau$, then $\bar\sigma_T(v) = \sigma(-\tau)$. As $\Delta(-\tau) = 0$, the triangle inequality gives
\begin{equation*}
|\Delta(v) - \Delta(u)| \le |\Delta(v) - \Delta(-\tau)| + |\Delta(u) - \Delta(-\tau)| = |\Delta(v)| + |\Delta(u)|,
\end{equation*}
and we bound both terms by $\frac{c}{4}\sigma(u)$. For the first, since $\sigma(t) \le e^{t}$ for all $t \in \mathbb{R}$ and $\sigma(t) \ge e^{t}/2$ for $t \le 0$,
\begin{equation*}
|\Delta(v)| = \sigma(-\tau) - \sigma(v) \le \sigma(-\tau) \le e^{-\tau} = \tfrac{c}{8}e^{-T} \le \tfrac{c}{4}\sigma(-T) \le \tfrac{c}{4}\sigma(u).
\end{equation*}
For the second, \Cref{loc:piecewise:linear_approximation_of_an_increasing_smooth_function.statement} at the pair $(u, -\tau)$ gives $|\Delta(u)| = |\Delta(u) - \Delta(-\tau)| \le \frac{c}{4}(\sigma(u) - \sigma(-\tau)) \le \frac{c}{4}\sigma(u)$. Hence $|\Delta(v) - \Delta(u)| \le \frac{c}{2}\sigma(u)$, while $\sigma(u) - \sigma(v) \ge \sigma(u) - \sigma(-\tau) \ge (1 - \frac{c}{4})\sigma(u) \ge \frac34\sigma(u)$, so
\begin{equation*}
|\Delta(v) - \Delta(u)| \le \tfrac{2c}{3}(\sigma(u) - \sigma(v)) \le c\,|\sigma(v) - \sigma(u)| .
\end{equation*}
If $v > \tau$, the same argument applies with $1 - \sigma$ in place of $\sigma$, since $1 - \sigma(t) = \sigma(-t)$.
\end{proof}
\begin{remark}
\label{loc:fixed:grid_approximation_of_the_sigmoid.the_tails_force_a_window}
At a fixed relative accuracy, the piecewise-linear approximation needs a number of pieces proportional to the length of the interval, and no piecewise-linear function with finitely many pieces approximates the sigmoid on all of $\mathbb{R}$ in this sense. This is why $\bar\sigma_T$ follows the increments of the sigmoid only from the points of the window $[-T, T]$, which holds the pre-activations of the signals away from saturation, and is constant on the tails.
\end{remark}
The scalar approximation carries over to the networks.
\begin{lemma}
\label{loc:fixed:grid_approximation_of_a_network_with_sigmoid_output.statement}
Let $H\colon \mathbb{R}^k \to \mathbb{R}^{k_{d-1}}$ be a network of depth $d-1$ as in \Cref{loc:feedforward_network_with_piecewise:linear_activations.statement} with hyperplane counts $M_1, \dots, M_{d-1}$, let $A_d:\mathbb{R}^{k_{d-1}} \to \mathbb{R}^n$ be affine, let $G\colon \mathbb{R}^k \to \mathbb{R}^n$, $G := \sigma \circ A_d \circ H$ with $\sigma\colon \mathbb{R}^n \to \mathbb{R}^n$ the sigmoid applied entrywise, let $T \in [0, \infty)$ and $c \in (0, 1)$, let $\bar\sigma_T \colon \mathbb{R}^n \to \mathbb{R}^n$ be the piecewise-linear approximation of the sigmoid from \Cref{loc:fixed:grid_approximation_of_the_sigmoid.statement} with accuracy parameter $c$, applied entrywise, and let $\bar G\colon \mathbb{R}^k \to \mathbb{R}^n$, $\bar G := \bar\sigma_T \circ A_d \circ H$. Then $\bar G$ is a network as in \Cref{loc:feedforward_network_with_piecewise:linear_activations.statement} with hyperplane counts $M_1, \dots, M_{d-1}$ and
\begin{equation*}
M_d \le n\Big(\frac{10}{c}\Big(T + \log\frac1c + 3\log 2\Big) + 2\Big),
\end{equation*}
and for all $z, w \in \mathbb{R}^k$ with $\|A_dH(w)\|_\infty \le T$,
\begin{equation*}
\|(G(z) - G(w)) - (\bar G(z) - \bar G(w))\|_2 \le c\,\|G(z) - G(w)\|_2 .
\end{equation*}
\end{lemma}
\begin{proof}[\hypertarget{loc:fixed:grid_approximation_of_a_network_with_sigmoid_output.proof}Proof of \Cref{loc:fixed:grid_approximation_of_a_network_with_sigmoid_output.statement}]\phantomsection\label{loc:fixed:grid_approximation_of_a_network_with_sigmoid_output.proof}

The bound of \Cref{loc:fixed:grid_approximation_of_the_sigmoid.statement} in coordinate $j$, at $(A_dH(z))_j$ with $u := (A_dH(w))_j \in [-T, T]$, squared and summed over $j$, is the accuracy claim. For the count, $\bar\sigma_T$ has one kink fewer than its pieces, and the hyperplanes $\{x_\nu = t\}$, for each entry $\nu$ and kink $t$, cover it on $\mathbb{R}^n$.
\end{proof}
\subsection{Recovery with Gaussian measurements}
\label{loc:body.smooth_output_activation.recovery_with_gaussian_measurements}
\begin{theorem}
\label{loc:gaussian_recovery_of_unsaturated_signals_without_truncation.statement}
Let $\delta \in (0, \frac12]$, let $H\colon \mathbb{R}^k \to \mathbb{R}^{k_{d-1}}$ be a network of depth $d-1$ as in \Cref{loc:feedforward_network_with_piecewise:linear_activations.statement} with hyperplane counts $M_1, \dots, M_{d-1}$. With $k_d := n$, let $b$ be the bottleneck layer of the widths $k_0, \dots, k_d$, so that $k_b = \min_{0 \le i \le d} k_i$. Let $A_d\colon \mathbb{R}^{k_{d-1}} \to \mathbb{R}^n$ be affine, and let $G\colon \mathbb{R}^k \to \mathbb{R}^n$ be the network $G := \sigma \circ A_d \circ H$, where $\sigma\colon \mathbb{R}^n \to \mathbb{R}^n$ is the sigmoid applied entrywise. Let $S_\delta := \operatorname{range}(G) \cap [\delta, 1 - \delta]^n$. Finally, let $A \in \mathbb{R}^{m \times n}$ be a matrix with independent $N(0, 1/m)$ entries. There are absolute constants $C_0, c_0 > 0$ such that if
\begin{equation*}
m \ge C_0\,k_b\Big(1 + \sum_{j=b+1}^{d-1}\log\Big(1 + \frac{M_j}{k_b}\Big) + \log\frac{n}{k_b} + \log\Big(1 + \log\frac1\delta\Big)\Big),
\end{equation*}
then with probability at least $1 - 3e^{-c_0 m}$ the following holds. For every signal $x_0 \in \mathbb{R}^n$, point $\bar x \in S_\delta$, noise $\xi \in \mathbb{R}^m$ and tolerance $\varepsilon \ge 0$, with measurements $y := Ax_0 + \xi \in \mathbb{R}^m$, and any $\hat x \in \operatorname{range}(G)$ such that $\|A\hat x - y\|_2^2 \le \inf_{x \in \operatorname{range}(G)}\|Ax - y\|_2^2 + \varepsilon$, we have
\begin{equation*}
\|\hat x - x_0\|_2 \le \|x_0 - \bar x\|_2 + 4\|A(x_0 - \bar x)\|_2 + 4\|\xi\|_2 + 2\sqrt\varepsilon .
\end{equation*}
\end{theorem}
\begin{proof}[\hypertarget{loc:gaussian_recovery_of_unsaturated_signals_without_truncation.proof}Proof of \Cref{loc:gaussian_recovery_of_unsaturated_signals_without_truncation.statement}]\phantomsection\label{loc:gaussian_recovery_of_unsaturated_signals_without_truncation.proof}

We first show that, with probability at least $1 - 3e^{-c_0 m}$, $A$ satisfies the $\frac12$-RIP on the set of differences $\operatorname{range}(G) - S_\delta$; the error bound then follows in a few lines. To get the RIP, we approximate $G$ by a piecewise-linear network $\bar G$, fixed before $A$ is drawn, whose range has few pieces, so that $A$ satisfies the RIP on its range, and we transfer the RIP from $\bar G$ to $G$.

We start with the approximation. Let $T := \log(1/\delta - 1)$, so that, as the sigmoid is increasing, $G(w) \in S_\delta$ exactly when $\|A_dH(w)\|_\infty \le T$. Let $c := \frac{1}{26}\sqrt{k_b/n}$, and let $\bar G := \bar\sigma_T \circ A_d \circ H$ be the network of \Cref{loc:fixed:grid_approximation_of_a_network_with_sigmoid_output.statement} with accuracy parameter $c$. As $T \le \log\frac1\delta$ and $\log\frac1c = \log 26 + \frac12\log\frac{n}{k_b}$, we have that $\bar G$ is a network as in \Cref{loc:feedforward_network_with_piecewise:linear_activations.statement} whose hyperplane counts $M_1, \dots, M_{d-1}$ match those of $G$, and whose last count satisfies $M_d \le C\,n\sqrt{n/k_b}\,\big(1 + \log\frac1\delta + \log\frac{n}{k_b}\big)$ for an absolute constant $C$.

We show that $A$ satisfies the RIP on the range of $\bar G$. With $N$ as in \Cref{loc:rip_on_the_range_of_a_piecewise:linear_network.statement}, $\log N = k_b\sum_{j=b+1}^{d}\log(1 + eM_j/k_b)$. Taking into account the bound on $M_d$, the sample complexity bound in the statement suffices for \Cref{loc:rip_on_the_range_of_a_piecewise:linear_network.statement}, which gives the $\frac13$-RIP on $\operatorname{range}(\bar G) - \operatorname{range}(\bar G)$ with probability at least $1 - 2e^{-c_0 m}$. Also, $\|A\|_{\mathrm{op}} \le 2 + \sqrt{n/m}$ with probability at least $1 - e^{-m/2}$ (\cite[Theorem 7.3.1]{vershyninHighDimensionalProbabilityIntroduction2018} and Gaussian concentration, \cite[Theorem 5.6]{boucheronConcentrationInequalitiesNonasymptotic2013}). Taking $c_0 \le \frac12$, by a union bound both events hold with probability at least $1 - 3e^{-c_0 m}$.

We now transfer the RIP from $\bar G$ to $G$. Conditioning on both events, as $m \ge k_b$,
\begin{equation*}
c\,(8 + 6\|A\|_{\mathrm{op}}) \le \tfrac{1}{26}\Big(20\sqrt{k_b/n} + 6\sqrt{k_b/m}\Big) \le 1,
\end{equation*}
which shows that $c$ is small enough to satisfy the conditions of \Cref{loc:transfer_of_the_restricted_isometry_between_two_networks.statement}. By \Cref{loc:fixed:grid_approximation_of_a_network_with_sigmoid_output.statement}, for every $z \in \mathbb{R}^k$ and every $w$ with $G(w) \in S_\delta$, the difference $G(z) - G(w)$ is within relative accuracy $c$ of $\bar G(z) - \bar G(w)$, so \Cref{loc:transfer_of_the_restricted_isometry_between_two_networks.statement} gives the $\frac12$-RIP on $\operatorname{range}(G) - S_\delta$.

Finally we bound the recovery error. Let $h := \hat x - \bar x \in \operatorname{range}(G) - S_\delta$. By the RIP and the choice of $\hat x$, which gives $\|A\hat x - y\|_2 \le \|A\bar x - y\|_2 + \sqrt\varepsilon$,
\begin{equation*}
\tfrac12\|h\|_2 \le \|Ah\|_2 \le \|A\hat x - y\|_2 + \|A\bar x - y\|_2 \le 2\|A(\bar x - x_0)\|_2 + 2\|\xi\|_2 + \sqrt\varepsilon .
\end{equation*}
The claim follows from $\|\hat x - x_0\|_2 \le \|h\|_2 + \|x_0 - \bar x\|_2$.
\end{proof}
In the ideal setting, where we have noiseless measurements ($\xi = 0$), no modeling error ($x_0 = \bar x \in S_\delta$) and an exact minimizer ($\varepsilon = 0$), we obtain exact recovery, $\hat x = x_0$. Also note that when $x_0$ lies off the range, $\|A(x_0 - \bar x)\|_2 \le 2\|x_0 - \bar x\|_2$ with high probability, provided $x_0$ and $\bar x$ are fixed before $A$ is drawn \cite{boraCompressedSensingUsing2017}.

Truncating the sigmoid at the saturation level removes the restriction to $S_\delta$.
\begin{corollary}
\label{loc:gaussian_recovery_for_a_truncated:sigmoid_output_from_the_unsaturated_theorem.statement}
Make the assumptions of \Cref{loc:gaussian_recovery_of_unsaturated_signals_without_truncation.statement}, but with the final activation of $G$ replaced by the sigmoid truncated at $\delta$, $\sigma_\delta\colon \mathbb{R}^n \to \mathbb{R}^n$, $\sigma_\delta(x)_\nu := \min\{\max\{\sigma(x)_\nu, \delta\}, 1 - \delta\}$. Then the guarantee of that theorem holds for every $\bar x \in \operatorname{range}(G)$.
\end{corollary}
\begin{proof}[\hypertarget{loc:gaussian_recovery_for_a_truncated:sigmoid_output_from_the_unsaturated_theorem.proof}Proof of \Cref{loc:gaussian_recovery_for_a_truncated:sigmoid_output_from_the_unsaturated_theorem.statement}]\phantomsection\label{loc:gaussian_recovery_for_a_truncated:sigmoid_output_from_the_unsaturated_theorem.proof}

We model the truncated sigmoid by adding an intermediate layer in front of the untruncated sigmoid, so that the modified network $G_\delta := \sigma_\delta \circ A_d \circ H$ becomes a network with the untruncated sigmoid whose whole range lies away from saturation. Let $T := \log(1/\delta - 1) \in [0, \infty)$, so that $\sigma(-T) = \delta$ and $\sigma(T) = 1 - \delta$, and let $\mathbf 1 \in \mathbb{R}^n$ be the all-ones vector. We build the layer before the sigmoid from ReLUs: let $A_d'\colon \mathbb{R}^{k_{d-1}} \to \mathbb{R}^{2n}$, $A_d'(x) := (A_d(x) + T\mathbf 1, A_d(x) - T\mathbf 1)$, let $\rho\colon \mathbb{R}^{2n} \to \mathbb{R}^{2n}$ be the ReLU applied entrywise, and let $A_{d+1}\colon \mathbb{R}^{2n} \to \mathbb{R}^n$, $A_{d+1}(z, z') := z - z' - T\mathbf 1$. Then $G_\delta = \sigma_\delta \circ A_d \circ H = \sigma \circ A_{d+1} \circ \rho \circ A_d' \circ H$. This adds one layer before the sigmoid, of width $2n$ with $2n$ hyperplanes, and changes nothing else in the network. Every entry of a point of $\operatorname{range}(G_\delta)$ lies in $[\delta, 1 - \delta]$, so \Cref{loc:gaussian_recovery_of_unsaturated_signals_without_truncation.statement}, at depth $d + 1$, applies to all of $\operatorname{range}(G_\delta)$. In its sample complexity bound the narrowest width stays $k_b$, and the added layer contributes at most one term $\log(1 + 2n/k_b)$, which is absorbed in the absolute constant $C_0$.
\end{proof}
\section*{Conclusion}
\label{loc:body.conclusion}
The main open problem arising from this work is uniform recovery without a saturation level: whether, for the untruncated sigmoid, the matrix $A$ satisfies, with high probability, the restricted isometry on all chords of $\operatorname{range}(G)$, with a sample complexity competitive with that for piecewise-linear networks. Finally, only the output activation of our networks is smooth; one avenue for future work would be to extend our results to networks with smooth hidden activations.
\section*{Acknowledgements}
\label{loc:body.acknowledgements}
The authors thank Sharvaj Kubal for many discussions and advice, and for his help in developing the piecewise-linear approximation of the sigmoid.

Y. Plan is partially supported by an NSERC Discovery Grant (GR009284) and an NSERC Discovery Accelerator Supplement (GR007657). O. Yilmaz was supported by an NSERC Discovery Grant (22R82411). O. Yilmaz also acknowledges support by the Pacific Institute for the Mathematical Sciences (PIMS) and the CNRS -- PIMS International Research Laboratory.

Artificial intelligence was used at all stages of writing this manuscript, including planning, proof writing, and polishing the manuscript. The core ideas were developed by the authors, and the manuscript was thoroughly manually edited and reviewed.
\section*{Data availability}
\label{loc:body.data_availability}
Data sharing is not applicable to this article as no datasets were generated or analysed during the current study.
\section*{Conflict of interest}
\label{loc:body.conflict_of_interest}
On behalf of all authors, the corresponding author states that there is no conflict of interest.
\bibliographystyle{plainnat}
\bibliography{bibliography}
\appendix
\section*{Appendix}\addcontentsline{toc}{section}{Appendix}
\begin{proof}[\hypertarget{loc:gaussian_recovery_of_unsaturated_signals_without_truncation.simplified_proof}Proof of \Cref{loc:gaussian_recovery_of_unsaturated_signals_without_truncation.simplified_statement}]\phantomsection\label{loc:gaussian_recovery_of_unsaturated_signals_without_truncation.simplified_proof}

The simplified statement is the special case of \Cref{loc:gaussian_recovery_of_unsaturated_signals_without_truncation.statement} for ReLU networks with $\bar x := x_0$, for which the first two terms of the error bound vanish. We only detail the points that are not immediate.

\emph{Hyperplane counts.} The ReLU on $\mathbb{R}^{k_j}$ is affine on each cell of the $k_j$ coordinate hyperplanes, so $M_j \le k_j$, and $k_b = k$ as the input layer is the narrowest.

\emph{Sample complexity.} The bracket in \Cref{loc:gaussian_recovery_of_unsaturated_signals_without_truncation.statement} is increasing in each $M_j$. As $k_j \ge k$, we have $\log(1 + k_j/k) \le \log 2 + \log(k_j/k)$, and $\sum_{j=1}^{d}\log(k_j/k) = d\,\ell$ with $\ell := \log(K/k)$. So the bracket is at most
\begin{equation*}
1 + (d-1)\log 2 + d\,\ell + \log\Big(1 + \log\frac1\delta\Big) \le 2\Big(d\,\ell + \log\Big(1 + \log\frac1\delta\Big)\Big),
\end{equation*}
where the requirement $K \ge e\,k$, that is $\ell \ge 1$, absorbs $1 + (d-1)\log 2 \le d$. So the simplified statement holds with $C := 2\,C_0$.
\end{proof}
\end{document}